\documentstyle[12pt]{article}
\begin{document}

\begin{titlepage}
\begin{flushright}
       {\bf UK/92-05}  \\
       Dec. 1992      \\
       hep-lat/9304011
\end{flushright}
\begin{center}

{\bf {\LARGE  Nucleon Axial Form Factor from Lattice QCD}}

\vspace{1.5cm}

{\bf    K.F. Liu, S.J. Dong, T. Draper, and J.M. Wu
 \footnote{ Present address: Institute of High
 Energy Physics, Beijing, China}}\\ [0.5em]
{\it  Dept. of Physics and Astronomy  \\
  Univ. of Kentucky, Lexington, KY 40506}
 \\[1em]
{\bf     W. Wilcox}\\ [0.5em]
  {\it Physics Dept., Baylor Univ., Waco, TX 76798}

\end{center}

\vspace{1cm}

\begin{abstract}

Results for the isovector axial form factors of the proton from a
lattice QCD calculation are presented for both point-split and local
currents. They are obtained on a quenched $16^{3} \times 24$ lattice at
$\beta= 6.0$ with Wilson fermions for a range of quark masses from
strange to charm. For each quark mass, we find that the axial form
factor falls off slower than the corresponding proton electric form
factor. Results extrapolated to the chiral limit show that the $q^2$
dependence of the axial form factor agrees reasonably well with
experiment. The axial coupling constant $g_A$ calculated for the local
and the point-split currents is about 6\% and 12\% smaller than the
experimental value respectively.  \\
\vskip\baselineskip

\end{abstract}

\vfill \end{titlepage}

As have electromagnetic form factors, the isovector axial form factor
$g_A(q^2)$ has been commonly used to constrain the construction of
models of the nucleon in order to incorporate PCAC and the
Goldberger-Treiman relation \cite{br79,anw83}. Similar to vector meson
dominance in electromagnetic form factors, the isovector axial form
factor seems to be quite sensitive to whether degrees of freedom in the
$A_1$  channel (e.g. $\rho\pi, \omega\pi\pi$ or $A_1$ itself) are
introduced in the effective theory \cite{ll87,mkw87,gk84}. In view of
the fact that the EM form factors and the magnetic moments of the
nucleon in recent lattice QCD calculations are within 10 to 15\% of the
experimental results \cite{dwl90,wdl92}, it is natural to extend the
study to the axial form factor and to determine to what extent the
experimental results can be reproduced in lattice QCD, subject to the
limitations of the quenched approximation, finite size effects and the
extrapolation to the chiral limit. Especially interesting is to check
the $q^2$ dependence of the axial vs. the electric form factors to see
if the experimental difference between them is borne out in the lattice
calculation; the $q^2$ dependence of their ratio should be fairly
independent of the systematic errors due to the present limitations of
the lattice calculation. In this paper, we study the axial form factor
in lattice QCD with Wilson fermions and compare it to the electric form
factor calculated previously and to experiment. We examine its
systematics as a function of the quark mass from strange to twice that
of the charm mass, and compare the axial coupling constant $g_A$ to
previous quenched calculations with different volumes
\cite{fpp82,wl88,glm89,gbb91}. Fixing $g_A$ to the non-relativistic
value of 5/3, we determine the finite lattice renormalization for heavy
quarks. Assuming axial dominance, we extract the $A_1 NN$ form factor.

Lattice gauge calculations have been carried out to study the
electromagnetic form factors of the pion\cite{dww89} and the
nucleon\cite{dwl90,wdl92}. The same sequential source technique (SST)
using the zero-momentum point nucleon interpolating field as the
secondary source is applied here to study the axial form factors
\cite{wdl92}. In choosing the final hadron as the secondary source, one
can sew the quark propagators together at the point where the current
couples in the three-point function. This has the advantage of being
able to study different currents at various momentum transfers.

The lattice two- and three-point functions that we calculate are the
following:
\begin{equation} \label{twopt}
G_{PP}^{\alpha\alpha}(t,\vec{p}) = \sum_{\vec{x}}e^{-i\vec{p}\cdot
\vec{x}}  \langle 0| T (\chi^\alpha(x) \bar{\chi}^\alpha(0) |0 \rangle ,
\end{equation}
\begin{equation} \label{threept}
G_{PAP}^{\alpha\beta}(t_f,\vec{p},t,\vec{q})=
\sum_{\vec{x}_f,\vec{x}} e^{-i\vec{p}\cdot\vec{x}_f
  +i\vec{q}\cdot\vec{x}} \langle 0| T(\chi^\alpha(x_f) A_\mu(x)
  \bar{\chi}^\beta(0)) |0 \rangle ,
\end{equation}
where $\chi^\alpha$ is the proton interpolating field and $A_{\mu}(x)$
is either the point-split axial vector current
\begin{equation} \label{psc}
A_\mu^{PS} \!=\! i 2\kappa [\bar{\psi}(x)\frac{1}{2}\gamma_\mu\gamma_5
U_\mu(x) \psi(x+\hat{\mu}) \!+\! \bar{\psi}(x+\hat{\mu})
\frac{1}{2}\gamma_\mu\gamma_5 U_\mu^\dagger (x)\psi(x)],
\end{equation}
or the local current
\begin{equation}   \label{loc}
A_\mu^{LOC} = i2 \kappa \bar{\psi}(x)\gamma_\mu\gamma_5 \psi(x).
\end{equation}
Phenomenologically, the axial vector current matrix element is written
as
\begin{equation}
 \langle \vec{p} s| A_\mu(0) |\vec{p}^{\;\prime} s^\prime\rangle
 = \bar{u}(\vec{p}, s) [i\gamma_\mu g_A(q^2)-q_\mu h_A(q^2)]\gamma_5
 u(\vec{p}^{\;\prime},s^\prime).
\end{equation}
It has been shown \cite{lwd91} that when $ t_f - t$ and $t >> a$, the
lattice spacing, the ratio of eqs.~(\ref{threept}) and~(\ref{twopt})
gives the following relations for the $A_3$ and $A_4$ matrix elements
\begin{eqnarray}
\frac{\Gamma^{\beta\alpha}G_{PA_3P}^{\alpha\beta}(t_f,\vec{0},t,
\vec{q})} {G_{PP}^{\alpha\alpha}(t_f,\vec{0})}
 & \longrightarrow& \frac{m+E_q}{2E_q}e^{-(E_q-m)t}
 [g_A(q^2)-\frac{q_3^2}{E_p+m}h_A(q^2)], \label{A3}  \\
\frac{\Gamma^{\beta\alpha}G_{PA_4P}^{\alpha\beta}(t_f,\vec{0},t,
\vec{q})} {G_{PP}^{\alpha\alpha}(t_f,\vec{0})}
 &\longrightarrow& \frac{q_3}{2E_q}e^{-(E_q-m)t}
 [g_A(q^2)+(m - E_q)h_A(q^2)],   \label{A4}
\end{eqnarray}
where
$\Gamma=\left(\begin{array}{cc}
                 \sigma_3  & 0 \\
                 0 & 0\end{array}\right)$.
When $\vec{q}=0$, eq.~(\ref{A3}) reduces to $g_A$, the coupling
constant. For finite momentum transfer, $g_A(q^2)$ can be obtained from
eq.~(\ref{A3}) upon eliminating the kinematic factor extracted from the
ratio of the two-point functions $G_{PP}^{\alpha\alpha}(t,\vec{q})/
G_{PP}^{\alpha\alpha}(t,0)$, and by setting $q_3$ to zero. The only
exception is when $\vec{q}\,^2 = 3(2\pi/La)^2$ where $q_3$ can not be
set to zero. In this case, we use both eqs.~(\ref{A3}) and~(\ref{A4})
to obtain $g_A (q^2)$ at this $\vec{q}\,^2$.

Our quenched gauge configurations were generated on a $16^3 \times 24$
lattice at $\beta = 6.0$. The gauge field was thermalized for 5000
sweeps from a cold start and 12 configurations separated by at least
1000 sweeps were used. Periodic boundary conditions were imposed on the
quark fields in the spatial directions. In the time direction, the
fixed (uncoupled) quark boundary condition was used. All quark
propagators were chosen to originate from lattice time slice 5; the
secondary nucleon source was fixed at time slice 20 (except for $\kappa
= 0.154$ where the quark propagators from time slice 3 to 22 are used).
We also averaged over the directions of equivalent lattice momenta in
each configuration; this reduces error bars.

The results presented here are for Wilson fermions with $\kappa =
0.154$, $0.152$, $0.148$, $0.140$, $0.133$, $0.120$ and $0.105$ so that
a range of quark masses from the strange to about twice the charm mass
is covered. To extract the lattice axial charge $g_A^L$, instead of
using the ratio in eq.~(\ref{A3}), we fit the three-point and two-point
functions to two exponentials in the form of $g_A^L fe^{-mt}$ and
$fe^{-mt}$ simultaneously using the data-covariance matrix to account
for the fact that they are measured on the same set of gauge
configurations~\cite{lll93}. The range of $t$ for the two-point
functions are chosen to overlap with the $t_f$ in the three-point
function. This is done except for the heavy masses (i.e. $\kappa =
0.120$ and $0.105$ for the point-split current and $0.140$ and $0.133$
in addition for the local current) where eq.~(\ref{A3}) is used. For
$q^2 \neq 0$, we used the combined ratios in eqs.~(\ref{A3})
and~(\ref{A4}) and $G_{PP}^{\alpha\alpha}(t,\vec{q})/
G_{PP}^{\alpha\alpha}(t,0)$ to extract $g_A^L(q^2)$, since we do not
have enough gauge configurations to warrant a simultaneous
fit~\cite{dgr91}. The errors are obtained through the jackknife in this
case.

In Table 1, we list the unrenormalized coupling constants $g_A^L =
g_A^L(q^2 = 0)$ for the point-split current (P-S.C.) and the local
current (L.C.) for different $\kappa$. We see that the results from the
P-S.C. are lower than those of the L.C. We will address this point
later.

\vskip .5cm
\begin{table}

\caption{Unrenormalized coupling constants $g_A^L$  calculated with the
point-split and the local currents  for different quark masses. Also
listed are the $\chi^2$ per degree  of freedom from the covariance
fit.}
\vskip\baselineskip

\begin{tabular}{|c|c|c|c|c|c|c|c|}
\hline
  $\kappa$ &   .154 & .152 & .148 & .140 & .133 & .120 & .105 \\
\hline
   $g_A^L$ (point-split)  & 1.28(17) & 1.26(6) & 1.24(4) & 1.11(1)
                        & 0.98(1) & 0.763(7) & 0.546(7)   \\
\hline
  $\chi^2 /N_{DF}$  & 0.28 & 0.03 & 0.77 & 0.43 & 1.3 & &   \\
\hline
\hline
   $g_A^L$ (local)        & 1.47(18)  & 1.43(5) & 1.40(5) & 1.25(1)
                        & 1.11(1) & 0.88(1) & 0.651(6)   \\
\hline
 $\chi^2/N_{DF}$    & 0.20 & 0.70 & 0.77 &  &  &  &  \\
\hline
\end{tabular}
\vskip .3cm
\end{table}

Plotted in Fig.~1 are the unrenormalized lattice isovector axial form
factors $g_A^L (q^2)$ of the proton as a function of the quark mass in
dimensionless units $m_q a = \ln [1 + 1/2(1/\kappa -1/\kappa_c)]$ for
different momentum transfers $\vec{q}\,^2 $ from $0$ to $4$ times
$(2\pi/La)^2$ for the point-split current. (N.B. $q^2 = (E - m_N)^2 -
\vec{q}\,^2$ for the four-momentum transfer squared.)  Also included
are earlier results for $g_A^L$ from smaller lattices ($10^3 \times
20$~\cite{wl88} and $12^3 \times 22$). Although error bars overlap, we
note that the finite volume effect is still appreciable for the light
quarks. A recent quenched calculation with the same spatial dimension
($16^3 \times 40$ with $\beta = 6.0$) \cite{gbb91} shows that $g_A^L$
at large $\kappa$ values of $0.154$ and $0.155$ for the local current
are at $1.36(4)$ and $1.39(14)$. Our result of $1.47(18)$ for $\kappa =
0.154$ is compatible with these and is also in agreement with the
earlier calculation with a renormalization improved Wilson
action~\cite{glm89}. The extrapolation of $g_A^L$ to the chiral limit
at $\kappa_c$ is carried out with the correlated fit to a linear
dependence on the quark mass $m_qa$ for $\kappa = 0.154$, $0.152$ and
$0.148$. The covariance matrix is calculated with the single
elimination jackknife error for $g_A^L$, itself calculated from the
simultaneous fit described above for different $\kappa$.

This extrapolation gives the unrenormalized $g_A^L$ at the chiral limit
to be $1.28(9)$ for the P-S.C. and $1.47(9)$ for the L.C. with
$\chi^2/N_{DF} = 0.01 $ for both cases. Using the finite lattice
renormalization $Z_L$ to be 0.86 for the P-S.C.~\cite{mm86} and 0.8 for
the L.C. \cite{gms89} as determined from current algebra matrix
elements, the continuum $g_A$ which equals $ Z_L g_A^L$ is 1.10(8) for
the P-S.C. and 1.18(7) for the local current. This is about 12\% and
6\% smaller than the experimental value of 1.254(6) respectively.

It was demonstrated in a previous study \cite{wl88} that the lattice
axial charge $g_A^L$ for the Wilson action is $(5/3)2\kappa$ for the
L.C. and  $\sim \kappa^2$ for the P-S.C. at the static limit. A quark
mass dependent factor $f(m_qa) = e^{m_qa} = 1 + 1/2 (1/\kappa -
1/\kappa_c)$ has been introduced to local scalar \cite{gbb91} and axial
currents \cite{glm89,gbb91} to smoothly interpolate between the limits
of $m_q \rightarrow 0$ where $f \rightarrow 1$ and $m_q \rightarrow
\infty$ where $f \rightarrow 1/(2\kappa)$ so that the static limit from
the Wilson action matches with the continuum results. A similar factor
is found in analyzing the Wilson action for non-relativistic quarks
\cite{lep91}. We multiply $g_A^L$ by this factor $e^{m_qa}$ for both
the L.C. and P-S.C. and plot them in Fig.~2 as a function of the quark
mass $m_qa$. It is seen that from $\kappa = 0.140$ on, the curve
flattens and the value is very close to the non-relativistic quark
model result of 5/3. We believe that this means the onset of the
non-relativistic limit is near $\kappa = 0.140$ and additional fine
tuning beyond the mean field finite lattice renormalization factor
$e^{m_qa}$ is needed to bring $g_A$ to be 5/3 for quark masses heavier
than $\kappa = 0.140$. In other words, the lattice field $\psi_L$ is
related to the continuum field $\psi_c$ through $\psi_c = (2\kappa
e^{m_qa}Z_A^L)^{1/2} \psi_L$. The factor $Z_A^L$ is listed in Table~2
for both the L.C. and the P-S.C. for the heavy quarks. To stress the
fact that these $\kappa$'s are {\it not\/} at the static limit, we
should mention that $g_A^L({\rm local})$ for $\kappa = 0.105$ is
$2\kappa(3.10)$ not the static value of $2\kappa (5/3)$. It is the
$Z_A^L e^{m_qa}$ factor that brings it to the non-relativistic value of
5/3. It is worthwhile pointing out that the finite lattice
renormalization can be done for heavy quarks in a gauge invariant way
without having to renormalize to quark properties which involves gauge
fixing. $Z_A^L$ for these quark masses is found to be very close to
unity for the L.C. This means that the $2\kappa e^{m_qa}$ is a
satisfactory ansatz for the renormalization. $Z_A^L$ for the P-S.C.
turns out to be 10 to 20\% larger. One would think that this is due to
the extra gauge-field link in the P-S.C. (eq.~(\ref{psc})) and that
replacing it with the mean field average $u_0$ which is between 0.8 and
0.9 at $\beta = 6$ would reconcile the difference between the
$Z_A^{L}$\,'s in these two currents. However, as we pointed out
earlier, the $g_A^L$ for the P-S.C. falls faster than that for the L.C.
by an extra $\kappa$ toward the static limit~\cite{wl88}. Hence, $Z_A^L
= Z_A^L(\kappa,g)$ is a function of both $\kappa$ and $g$. It can not
be related to the mean field alone.

\begin{table}
\caption{Renormalization factor $Z_A^L$ for the heavy quarks}
\vskip\baselineskip
\begin{center}
\begin{tabular}{|c|c|c|c|c|c|c|c|}
\hline
  $\kappa$ &    .140 & .133 & .120 & .105 \\
\hline
   $Z_A^L$ (local)  & 0.98(1) & 0.97(1) & 0.97(1) & 1.00(1) \\
\hline
   $Z_A^L$ (point-split)  & 1.10(1)  & 1.09(1) & 1.11(1) & 1.19(2) \\
\hline
\end{tabular}
\end{center}
\vskip .3cm
\end{table}

For $q^2 \neq 0$, we extrapolate the form factor to the chiral limit
with the procedure described in Ref.~\cite{gbb91}. The error of the
linear extrapolation is calculated from the single elimination
jackknife method. For each jackknife sample, the fit is carried out
with the covariance matrix to calculate $\chi^2/N_{DF}$. The average
$\chi^2/N_{DF}$ turns out to be about 1 or less.

In Fig. 3, we plot the axial form factor extrapolated to the chiral
limit for the L.C. and the P-S.C.
in comparison with the experimental result. In doing
so, we have used the calculated nucleon mass to set the scale for the
momentum as was done in ref.~\cite{wdl92}. For comparison, we also show
the calculated electric form factor $G_E$ extrapolated to the chiral
limit~\cite{wdl92} (without the covariance matrix) and the
corresponding experimental results. The experimental $g_A(q^2)$ has
been measured in neutrino-neutron scattering and pion
electroproduction. The neutrino data gives a good fit in the dipole
form up to $q^{2} = 3 {\rm GeV}^2/c^2$ \cite{baker}, i.e. $g_{A} (q^2)
= g_{A} (0)/ (1 - q^2/M_{A}^2)^2, $ with the axial vector coupling
constant $g_{A}(0) =1.254\pm 0.006$ and $M_{A} = 1.032\pm 0.036 {\rm
GeV}$ (world average). Our fit of the axial form factor to the dipole
form yields $M_{A} = 1.03 \pm 0.05  {\rm GeV}$ for the P-S.C. and $M_A
= 1.03 \pm 0.03 {\rm GeV}$ for the L.C. which are very close to the
experimental dipole mass. Similarly, the fitted dipole mass for $G_E$
is $M_{E} = 0.77 \pm 0.03 {\rm GeV}$ which is close to the experimental
dipole mass of $0.828\pm 0.006 {\rm GeV}$. We should quickly point out
that the favorable agreement with the experiments to a few per cent
level for the dipole mass could be fortuitous at this stage since we
have not adequately assessed the systematic errors due to the finite
size effect, the dynamical fermion effect, etc. Especially, the
isovector part of the charge radius of the nucleon has a chiral $\ln
m_{\pi}^2$ divergence while approaching the chiral limit \cite{bz72}.
This correction to the lattice extrapolation to the physical pion mass
can increase the charge radius by a significant amount \cite{lc93}.
However, unlike the isovector vector current which can couple to two
$\pi$'s leading to a pion loop, the isovector axial current couples to
three $\pi$'s. Hence, it is not subjected to the chiral $\ln m_{\pi}^2$
correction as alluded to for the charge form factor. Notwithstanding
these corrections (e.g. finite volume, dynamical fermion, chiral
limit), we want to stress the fact that the $q^2$ dependence of the
axial form factor shows that it is harder (i.e. falls off slower) than
the corresponding $G_E(q^2)$ at each quark mass we studied. This
feature, shown at the chiral limit in Fig. 3, is consistent with the
experimental data (i.e. the dipole mass of 1.03 {\rm GeV} for the axial
case is higher than the 0.828 {\rm GeV} for $G_E$). This feature, we
believe, is likely to survive the various lattice and chiral
corrections.

As far as the $q^2$ dependence is concerned, we find that the falloff
is faster as the quark mass decreases. This is evidenced in Fig. 4(a)
where $g_A^L(q^2)/g_A^L(0)$ for the P-S.C. is plotted for
different quark
masses. This is consistent with the fact that the associated meson
cloud, specifically the $A_1$, will have a larger Compton wavelength as
the meson masses decrease with the quark mass. In view of the success
of the vector dominance and the Goldberger-Treiman relation, the idea
of axial dominance for the isovector axial form factor seems to work
well phenomenologically~\cite{gk84}. With the axial dominance
assumption in mind, $g_A (q^2)$ can be written as
\begin{equation}  \label{A_1NN}
g_A(q^2) \sim g_{A_1 NN} (q^2) /(1 - q^2/m_{A_1}^2)
\end{equation}
where $g_{A_1 NN}(q^2)$ is the $A_1 NN$ form factor. This is calculated
by $g_A^L(q^2) (1 - q^2/m_{A_1}^2)$ at $\kappa= 0.148$, $0.152$, and
$0.154$ and extrapolated to the chiral limit. The results for the P-S.C.
normalized at $q^2 = 0$, are plotted in Fig. 4(b). We see that the
$q^2$ falloff is now less sensitive to the quark mass than the case of
$g_A^L(q^2)$. Finally we fit the $g_{A_1 NN}(q^2)$ obtained this way
with a monopole and found a monopole mass of $1.08 \pm 0.06 {\rm GeV}$
for the P-S.C. and $1.01 \pm 0.04 {\rm GeV}$ for the L.C.

To conclude, we have calculated the isovector axial form factor of the
nucleon for quark masses from strange to two times the charm mass.
Albeit it is a quenched approximation, we find that $g_A$, extrapolated
to the chiral limit, is about 6\% smaller than in experiment for the
local current and about 12\% smaller for the point-split current. It is
important to include the $e^{m_qa}$ correction factor for heavy quarks.
The additional finite lattice correction for the local and point-split
axial currents has been determined for heavy quarks in a gauge
invariant way. We note that $g_A(q^2)$ is harder than $G_E(q^2)$ of the
proton which is in agreement with experiment. This finding is most
likely to be preserved when finite volume, chiral, and dynamical
fermion effects are included. Assuming axial vector dominance, we have
extracted the $A_1 NN$ form factor. For future studies, it is essential
to improve the calculation by expanding the volume and incorporating
dynamical fermions.

This work is partially supported by the DOE Grand Challenge Award, DOE
Grant No. DE-FG05-84ER40154 and NSF Grants No. STI-9108764 and PHY-
9203306. We would like to thank R.M. Woloshyn for providing us with his
unpublished results on the $12^{3}\times22$ lattice.

\vfill

\noindent
{\bf Figure Captions}
\vskip\baselineskip
\noindent
Fig.~1 \hspace{1ex} The unrenormalized isovector axial form factor
$g_A^L(q^2)$ for the point-split current as a function of the quark
mass $m_qa$. The top curve is for $q^2 = 0$, the rest are for
$\vec{q}\,^2 $ from 1 to 4 times of $(2\pi/La)^2$ in descending order.
\vskip\baselineskip
\noindent
Fig.~2 \hspace{1ex} The lattice $g_A^L$ multiplied by the finite
lattice correction factor $e^{m_qa}$ for the point-split and the local
currents as a function of the quark mass.
\vskip\baselineskip
\noindent
Fig.~3 \hspace{1ex}  The calculated $g_A(q^2)$ for the L.C. and
$G_E(q^2)$ of the proton as a function of $-q^2$. The dashed line is
the fit to the experimental $g_A(q^2)$ with a dipole mass of $1.03 {\rm
GeV}$. The solid curve is the same for the experimental $G_E(q^2)$ with
a dipole mass of $0.828 {\rm GeV}$.
\vskip\baselineskip
\noindent
Fig.~4 \hspace{1ex} (a) $g_A^L(q^2)/g_A^L$ for the P-S.C.
for different quark masses as a function of $-q^2a^2$. \\
\hspace{1.3cm} (b) $g_{A_1 NN}(q^2)/g_{A_1 NN}(0)$ defined in
eq.~(\ref{A_1NN}) from the P-S.C. for several quark masses.

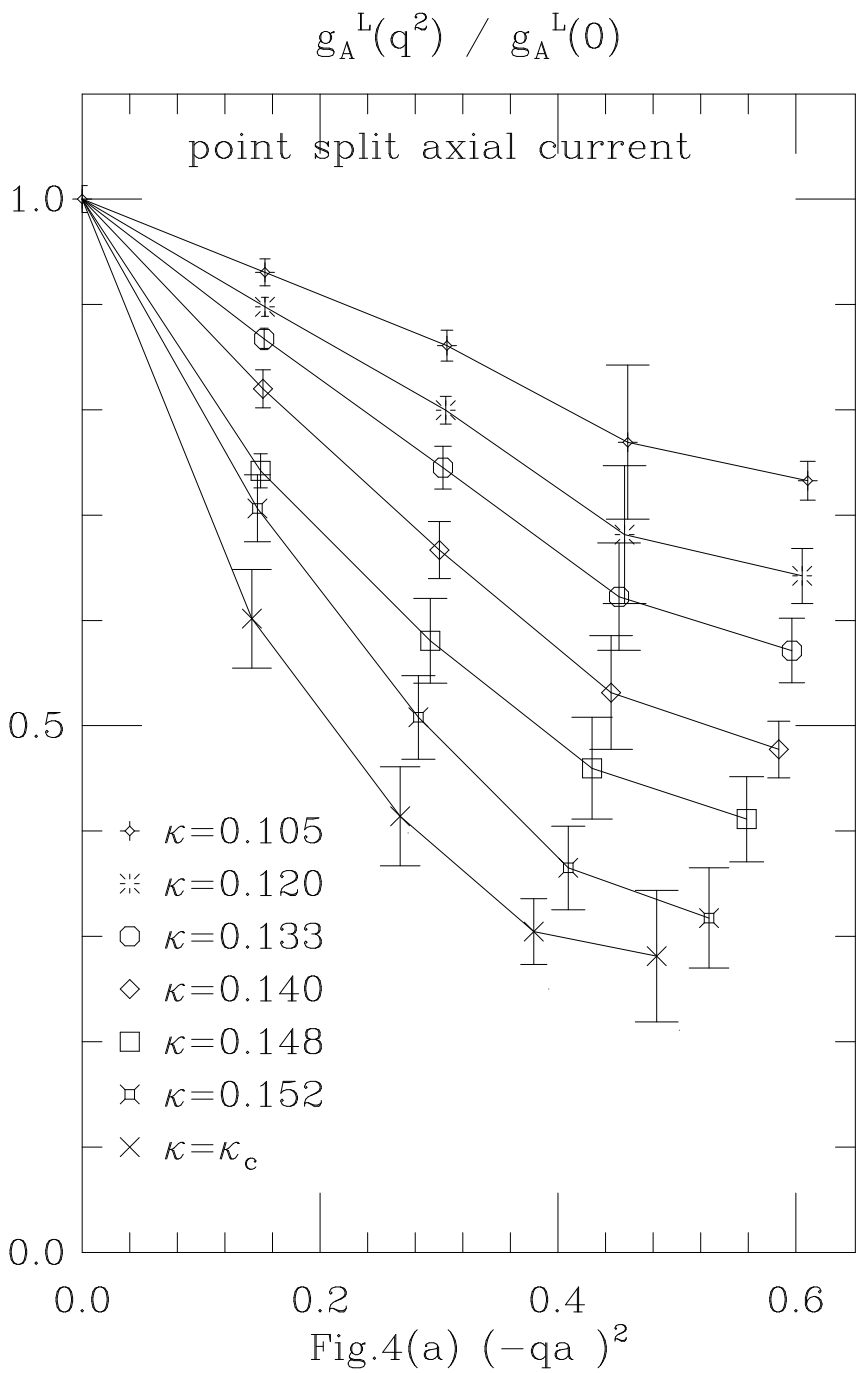
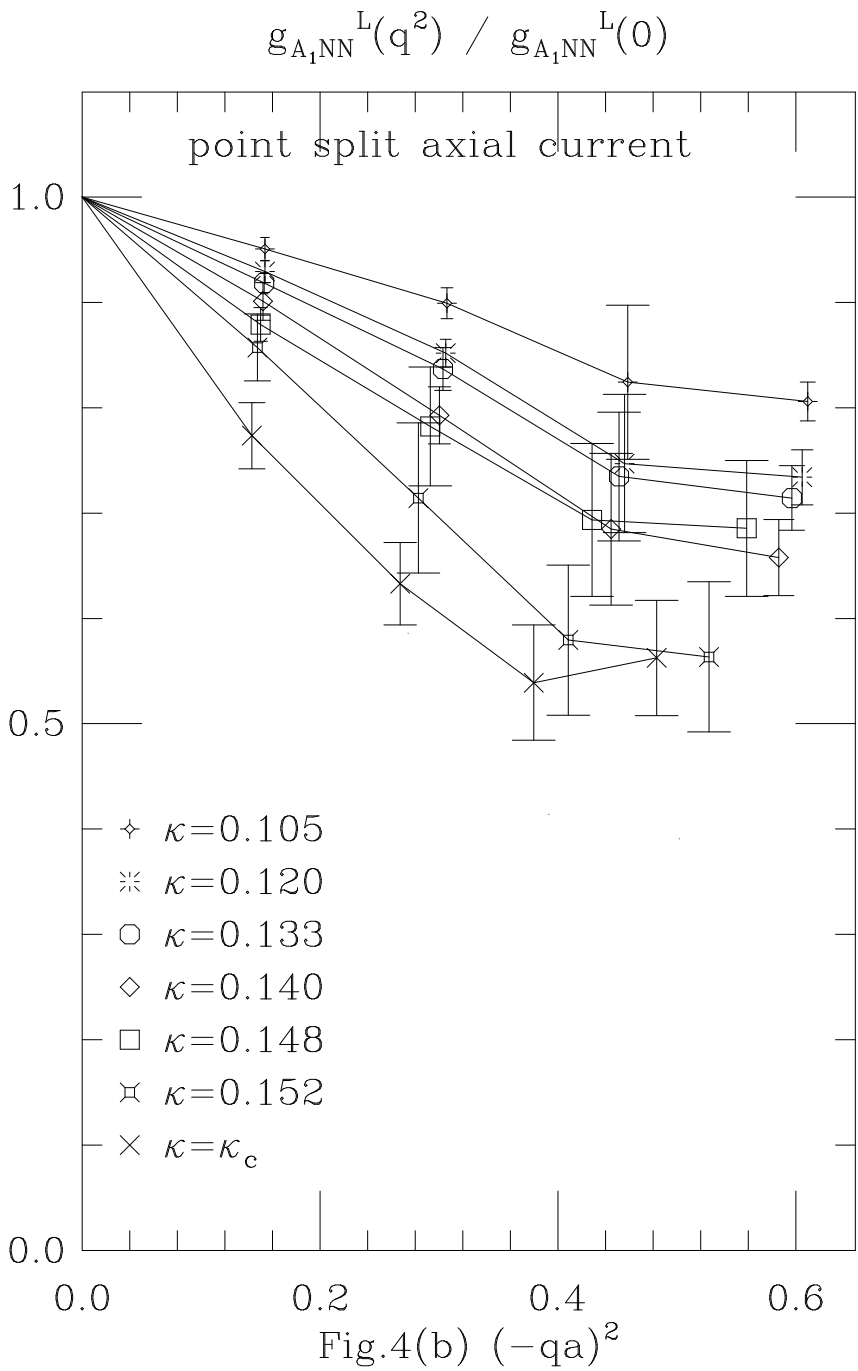
\end{document}